# Open Datasets for AI-Enabled Radio Resource Control in Non-Terrestrial Networks

Husnain Shahid*, Miguel Ángel Vázquez*, Laurent Reynaud†, Fanny Parzysz†, Musbah Shaat*
*Centre Tecnològic de Telecomunicacions de Catalunya (CTTC/CERCA), Castelldefels, Barcelona, Spain
† Orange Labs Network, Lannion, France

*Abstract*—By effectively implementing the strategies for resource allocation, the capabilities, and reliability of non-terrestrial networks (NTN) can be enhanced. This leads to enhance spectrum utilization performance while minimizing the unmet system capacity, meeting quality of service (QoS) requirements and overall system optimization. In turn, a wide range of applications and services in various domains can be supported. However, allocating resources in a multi-constellation system with heterogeneous satellite links and highly dynamic user traffic demand pose challenges in ensuring sufficient and fair resource distribution. To mitigate these complexities and minimize the overhead, there is a growing shift towards utilizing artificial intelligence (AI) for its ability to handle such problems effectively. This calls for the development of an intelligent decision-making controller using AI to efficiently manage resources in this complex environment. In this context, real-world open datasets play a pivotal role in the development of AI models addressing radio control optimization problems. As a matter of fact, acquiring suitable datasets can be arduous. Therefore, this paper identifies pertinent real-world open datasets representing realistic traffic pattern, network performances and demand for fixed and dynamic user terminals, enabling a variety of uses cases. The aim of gathering and publishing the information of these datasets are to inspire and assist the research community in crafting the advance resource management solutions. In a nutshell, this paper establishes a solid foundation of commercially accessible data, with the potential to set benchmarks and accelerate the resolution of resource allocation optimization challenges.

*Keywords*—Open Datasets, Non-terrestrial Networks (NTN), Resource Management, Artificial Intelligence.

## I. INTRODUCTION

In the realm of non-terrestrial networks (NTN), the efforts on efficiently utilizing the satellite resources are becoming a priority and requiring an attention, both in online and offline scenarios. This necessity becomes particularly pronounced in the context of multi-orbital and multi-band satellite integrated networks as it introduces complexities due to the dynamic nature of the satellite links and non-uniform user traffic demand that spans diverse application areas, including maritime, aeronautical, and land transport, among others [1].

To emphasize the importance of radio resource management (RRM) problem, significant efforts have been made to optimize the offered capacity of the whole system, taking into account the heterogeneous complexities and by leveraging the recently launched flexible payloads reconfiguration capabilities. In the sense of flexible satellite payload capabilities for (RRM), [2] optimized the capacity management of the system using Simulated Annealing (SA) algorithm to minimize the objective function that [3] delves into global resource management by integrating the Quality of Service (QoS) metric and user channel conditions, revealing valuable insights. Meanwhile, [4] introduces a joint optimization tool which simultaneously optimizes the beam width and bandwidth allocation to effectively cater the varying traffic demands. Nevertheless, these classical methods exhibit limited efficacy in dealing with extensive system-related constraints and the management of resources at a large scale and thus leading to sub-optimal solutions. Responsively, AI-driven approaches hold significant potential as key contributors to resource optimization process. In this context, [5] proposes a Deep Reinforcement Learning (DRL) algorithm to optimize the unmet system demand and power consumption using proximal policy optimization (PPO). The work in [6] mitigates the problem of inter-beam interference in multi-beam satellite system by scheduling the users and allocating bandwidth and power resources intelligently, mainly employing DRL techniques. In another study, the optimal long term capacity allocation plan in a highly complex three-layer heterogeneous satellite network is studied which uses the reinforcement learning as a framework to optimize the capacity allocation but in the realm of synthetic data [7].

However, it is widely acknowledged that the efficacy of AI paradigm heavily relies on the availability of large and diverse data to effectively learn from the variable environment and provide efficient solutions for resource management. In light of this recognition, the core objective of this paper centers on the identification of real-world open datasets that offer practical representations and are well-suited to be applied for resource management optimization tasks across various use cases. By identifying these datasets, the aim is to gain more practical insights into the realistic and non-uniform behavior of users' resources demand. This, in turn, will enable the development of AI-enabled intelligent decision controller tailored for NTN in real-world scenarios. This AI-based controller can leverage the realistic properties of the datasets and optimize the allocation of resources while serving the maximum number of users simultaneously.

Taking the objectives into account, the paper is organized as follows: Section II focuses on identifying the potential use cases for NTN and the corresponding open datasets, discussing the dataset attributes and the guidelines about utilization of these datasets for resource allocation task. Section III presents realistic user traffic patterns (UTPs) for understanding per user resource requirements. Section IV provides an example of one of the datasets at small scale to elaborate how the available information can be used to perform, for instance, beamforming to serve the user needs. Finally, the conclusion is drawn in Section V.

## II. POTENTIAL USE CASES FOR NTNs

The exploration of potential use cases and the understanding of the sectors that would benefit from NTN communication system are of paramount importance. In this regard, this section aims to highlight promising use cases while concurrently identifying potential datasets that could be used to enhance RRM more effectively. To gather relevant datasets, a comprehensive investigation is carried out across a variety of online platforms encompassing open data portals, online databases, and specialized dataset repositories. Furthermore, the data licensing and documentation are carefully examined to ensure the reliability and credibility of the datasets. This rigorous evaluation involves careful examination of the licensing agreements, terms of use, and documentation accompanying the datasets to ensure their suitability for research and analysis purposes. The prospective use cases are as follows.

### A. Maritime Use Case

As is widely recognized, NTN possesses the remarkable ability to extend connectivity across expansive maritime territories and overcome limitations of terrestrial networks. In this essence, non-terrestrial networks hold a significant potential for transforming the maritime industry, for instance enabling the seamless connectivity for the passenger yachts and facilitating the reliable communication among vessels, offshores platforms, maritime surveillance, and ultimately enhancing the rescue operations efficiency and meeting the safety requirements effectively. Therefore, by leveraging the NTN, this industry can significantly benefit from seamless connectivity and communication capabilities. One of the preeminent datasets for optimizing the resources in maritime, a weeklong dataset with the resolution of one hour, is acquired from a reputable data provider (VesselFinder VT Explorer) [8]. This dataset consists of realistic marine traffic information updated every one hour obtained from Automatic Identification System (AIS) over the entire Mediterranean Sea. The data attributes are shown in Figure 1.

While Figure 2 illustrates a snapshot of the dataset represented as a vessel density grid map. This grid map is a function of longitude and latitude where each grid encompassing an area of 100 km x 100 km using haversine distance. The haversine distance can be calculated using the following equations.

$$a = \sin^2\left(\frac{\Delta\phi}{2}\right) + \cos\phi_1 \cdot \cos\phi_2 \cdot \sin^2\left(\frac{\Delta\lambda}{2}\right) \quad (1)$$

$$c = 2 \cdot \arctan 2\left(\sqrt{a}, \sqrt{1-a}\right) \quad (2)$$

$$d = R \cdot c \quad (3)$$

where $R$ is the earth radius i.e., 6371 km, $\Delta\phi$ and $\Delta\lambda$ are the absolute difference between the latitudes ($\phi_1, \phi_2$) and longitudes ($\lambda_1, \lambda_2$) respectively and $d$ is haversine distance.

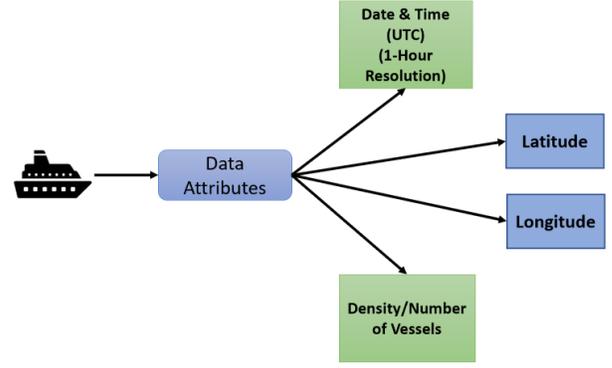

Fig. 1. The attributes of marine traffic dataset provided as an open dataset.

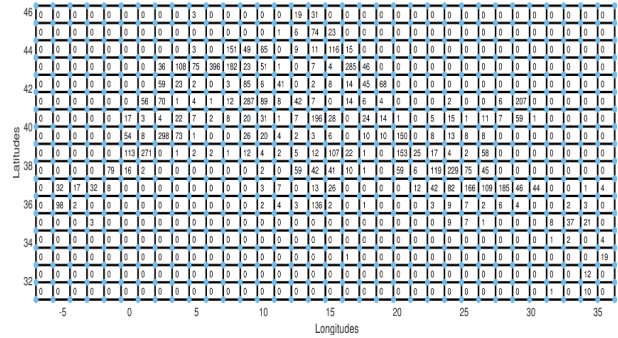

Fig. 2. The snapshot of marine traffic density of an hour duration.

*Guidelines for Maritime Dataset:* To construct this grip map, the latitudes of Mediterranean Sea are evenly spaced with the factor of 0.9 which is equivalent to the haversine distance of 100 km w.r.t latitudes while keeping the longitude constant. This process determines the vertical lines of a single grid as shown in Figure 2. Subsequently, the longitudes of Mediterranean Sea are evenly spaced with the factor of 1.2 which is equivalent to the haversine distance of 100 km w.r.t longitudes, while maintaining the latitudes constant, resulting in horizontal lines of the grid. By combining these vertical and horizontal lines, a complete grid map is formed. The numerical value within each grid cell represents the count of vessels observed within the corresponding 100 km x 100 km area. This grid-based representation allows for comprehensive analysis and visualization of maritime traffic variation within the dataset, offering insights into spatio-temporal distribution across different geological regions of Mediterranean sea.

The offered maritime dataset is compatible to imply a various AI based resource management algorithms after splitting the data into training and testing sets, for instance, traffic forecasting using deep learning which offers the opportunity to pre-allocate the resources [10], resource scheduling to take full advantage of limited resources [11], and adaptive beamforming for specific users based on their throughput demand etc. [4], aggregating per terminal demand provided in Section III.

The given repository in this paper provides open access to this dataset, which is stored in a .mat file format. Each hour of the dataset contains a matrix representing the traffic distribution values. The maritime traffic flow dataset (processed files) is accompanied by permissive license that allows users to access, commercially use, modify, and distribute the data freely. However, it is essential to give credits to the creators of the dataset (CTTC) as a sign of acknowledgement.

TABLE I
THE ATTRIBUTES OF GLOBAL FIXED AND MOBILE NETWORK PERFORMANCE DATA MAP BY OOKLA [9].

| No. | Key Performance Indicators (KPIs) | Description | Data type |
|---|---|---|---|
| 1 | Tile | The coordinates (longitude and latitude) of the tile | Numeric |
| 2 | avg_d_kbps | The average download speed in kbps* | Integer |
| 3 | avg_u_kbps | The average upload speed in kilobits per second | Integer |
| 4 | avg_lat_ms | The average latency measured in ms* | Integer |
| 5 | avg_lat_d_ms | The average latency during the download phase measured in ms | Integer |
| 6 | avg_lat_up_ms | The average latency during the upload phase measured in ms | Integer |
| 7 | Tests | The number of tests taken in the tile | Integer |
| 8 | Devices | The number of unique devices contributing tests in the tile | Integer |
| 9 | Quadkey | The quadkey representing the tile | String |

*B. PPDR Use Case*

Taking the into account the general term of public protection and disaster relief (PPDR) or temporary event, nomadic traffic can be one of the examples of PPDR event that involves the movement of travelers and vehicle motion across different locations, for instance, remote or underserved environments such as mountainous and isolated islands where terrestrial communication infrastructure may be insufficient or unavailable at all. Therefore, satellite communications merge as a valuable support in the realm of reliable and ubiquitous connectivity and seamless communication for nomadic travelers to access real time navigation system, location tracking, regardless of their location. This becomes particularly crucial in emergency situations where reliable communication is essential to enhance safety, efficiency, and convenience for the travelers on the move.

Obtaining the dataset to enable PPDR can be challenging due to the difficulties in identifying the areas with limited terrestrial communication infrastructure where the network performance is not up to the mark to serve even for temporary events. Moreover, personal data privacy and security laws impose strict obligations on the collection, storage and publication of any data related to people in the EU, including their mobility and geolocalization, in complying with the general data protection regulation (GDPR) or other national regulations (e.g., issued by the CNIL in France).

In this context, an option involves conducting surveys and connectivity tests to generate the dataset from scratch. However, this approach can be resource-intensive and time-consuming if adopted from scratch. Alternatively, utilizing an existing dataset is another possibility but it is often challenging to find datasets with open licenses. Fortunately, Ookla´s Open data facilitates with a trusted repository of worldwide network performances that offers accessible and modifiable data with open license and thereby adaptable for resource management tasks [9] [12]. Furthermore, the Urban Data Platform Plus (UDPP) by European Commission put additional efforts to Ookla´s data, by offering valuable insights into broadband speed for both fixed and mobile networks at the municipalities level across the European region [13] [14].

*Guidelines for Ookla Dataset:* This Ookla's open dataset facilitates with the global fixed broadband and mobile network performance in the form of tiles. The fixed broadband refers to measurements taken from mobile devices with non-cellular type connection i.e., Wi-Fi or Ethernet, whereas the mobile network measurements are taken from mobile devices with cellular type connection, for instance, 4G LTE and 5G NR. The dataset spans from the period of Q1 2019 to the recently complete quarter Q2 2023, being updated every quarter year (three months) and generated based on device participation and tests repetitions. The tests are performed at random hours with several devices, or a single device performs multiple tests in some cases. The measurements by all devices and tests are then aggregated to get average value in each tile. Since there is no information available about time at which the tests are being performed, therefore an educated guess is to trust these average values as it is. The attributes are shown in Table I.

However, the relevance of this data to resource allocation for NTN lies in the idea that, there are many tiles around the globe where the network performance is inadequate. Therefore, for the regions with poor network performance and no coverage, the implementation of NTN becomes necessary to fill the gaps to improve connectivity and thereby the management of resources based on requirements. The data provided by the dataset possess sufficient capability to be used as input for AI-enabled algorithms for instance several reinforcement learning algorithms [15] to optimize the network throughput in accordance with the identified insufficient network performance regions within this dataset and also the K-means (unsupervised machine learning) algorithm [4] useful for traffic demand based adaptive beamforming to serve the only areas where network performance is inadequate, as given in the example to be discussed in Section IV.

*C. Direct-to-Smartphone Use Case*

A direct-to-smartphone use case within the realm of NTNs is of great significance, particularly in the situations when the connectivity is needed through satellite networks for various temporary and permanent purposes. Moreover, the need for NTN becomes crucial for this use case when considering the provision of connections in remote villages that lack the overall terrestrial infrastructure. In such a scenario, NTN enables users to access the internet, make voice calls and exchange text messages. Another prominent scenario is during emergencies, for instance, natural disasters where the terrestrial network may be severely impaired. This enables the relief for the affected individuals to seek emergency assistance and communicate about their situation to authorities. Similarly, it empowers the authorities in the affected areas to communicate and expedite the rescue operations in the absence of terrestrial facilities.

*Guidelines Direct-to-Smartphone Dataset:* Based on the specified use case, the task of finding the suitable and readily available dataset for direct-to-handheld is deemed extremely challenging. However, one approach could be to merge the multiple datasets to extract a more accurate representation of the requirements considering both temporary (disaster situations) and permanent (remote and unserved areas) needs. In this essence, the strategic integration of the Ookla dataset, as [16], presents an opportunity to derive insights about the estimated requirements of NTN system design to better serve the direct-to-handheld use case.

The acquisition of high-resolution population density map involves employing state of the art computer vision techniques on high resolution (50 cm per pixel) satellite imagery data integrated with the publicly available census data. The idea behind building the map for population density involves using the convolutional neural network (CNN) to identify the buildings from publicly accessible mapping services such as OpenStreetMap (OSM). The population density map exhibits impressive accuracy level, providing data with 30 x 30 meter grid resolution, making it reliable and authentic dataset for utilization.

Since, Meta dataset offers statistical information regarding the population count at geocoordinates $X$ while the geocoordinates $Y$ of Ookla dataset is a subset of geocoordinates $X$, i.e. ($Y \subseteq X$) on which it presents the fixed and mobile network performance. By leveraging the population density from the Meta's dataset and examining the mobile network performance for underserved and remote areas and then locating the areas that are below a specific minimum threshold in terms of network performance, valuable insights can be obtained regarding the required resources in that specific area, for instance bandwidth required for a system to meet the requested throughput and resource scheduling in the presence of limited resources etc., that NTN should allocate to meet the unserved demand based on population density. The general scheme is depicted in Figure 3.

## III. USER TRAFFIC PATTERNS

So far, the datasets are extensively analyzed for their relevance to prominent use cases, providing valuable insights to identify where NTN resources are critically required. However, to accurately manage the resources, it is essential to have an important component, i.e. per user terminal demand. Although, Ookla and Meta AI datasets can help identifying the underserved areas and user terminals density, it lacks details about the per user terminal demand in those remote and underserved areas where terrestrial networks might be accessible with limited capacity, or it is not accessible at all.

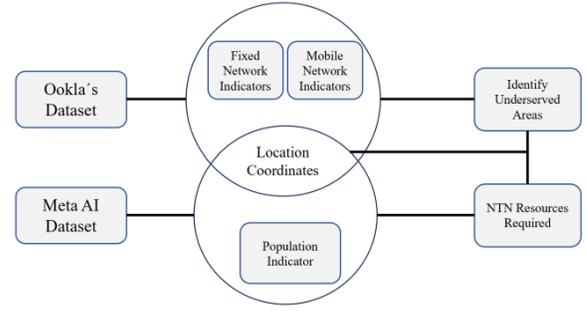

Fig. 3. The integrated data guidelines to identify the resources required for underserved/remote areas.

In this essence and regarding the cases particularly related to nomadic traffic and direct-to-smartphone, this section provides realistic traffic pattern for the purpose of integration with the provided (Ookla and Meta AI) datasets to complete the input information typically required for designing the AI-enabled controller. The data refers to user demand as user traffic profile (UTP), is a collection of key performance indicators (KPIs).

As an example, Table II outlines this set of KPIs for the direct-to-smartphone or nomadic case. In this regard, a UTP encompasses KPIs such as [17], among other performance targets. Their values are derived from the analysis of several direct-to-smartphone subcases. For instance, the UTP outlined in Table 2 relates to a use case whereby users are trekking in areas without continuous terrestrial coverage and wish to continue using their audio chat and video streaming services, both in outdoor and indoor conditions, with different quality of experience. This specific subcase notably highlights high user uplink (UL) requirements, which can be seen via the high values on the uplink experienced user data rate and activity factor. It completes a more general case in which the activity factor downlink (DL) is significantly higher than the UL.

Those UTPs are associated with specific areas (e.g., general UTPs target locations not covered by terrestrial networks, whereas the UTP illustrated by Table II is more specific and relates to mountainous and other remote areas where users are susceptible to require a service continuity for the defined trekking applicative needs).

The description of KPIs is detailed as follows [17],

- **Experienced User Data Rate:** Represents the minimum data rate needed to achieve a target Quality of Experience (QoE).
- **Activity Factor:** Ratio of simultaneous active UEs to the total number of UEs.
- **Monthly User Traffic:** Monthly User Traffic is expressed in GB/user/month for both the downlink and uplink.
- **Busy Hour Usage Rate:** Represented as,
  Busy Hour Usage Rate DL = Experienced Data Rate DL * Activity Factor DL

TABLE II
THE USER TRAFFIC PROFILE FOR DIRECT-TO-SMARTPHONE USE CASE

| Direct-to-Smartphone | UE type | Experienced user Data Rate | | Activity Factor | | Monthly User Traffic | | Busy Hour Usage Rate | | Latency | Radio Link Availability | Reliability | Mean User Density | Area Traffic Capacity | | User Distribution | UE Speed |
|---|---|---|---|---|---|---|---|---|---|---|---|---|---|---|---|---|---|
| | | DL | UL | DL | UL | DL | UL | DL | UL | | | | | DL | UL | | |
| | (Text) | Mbps | | % | | GB/Month | | Kbps | | ms | % | % | user/km$^2$ | Kbps/km$^2$ | | (Text) | km/h |
| Outdoor | Handheld | 2 | 10 | 5.0 | 5.0 | 167 | 33 | 100 | 500 | < 30 | 99.9 | 99.999 | 5 | 500 | 2500 | UEs distributed on trails | 20 |
| Light indoor | Handheld | 0.01 | 0.01 | 5.0 | 5.0 | | | 0.5 | 0.5 | < 50 | 99.9 | 99.99 | 5 | 2.5 | 2.5 | | |

Busy Hour Usage Rate UL = Experienced Data Rate UL * Activity Factor UL

- **Mean User Density:** Represents the average user density over a defined area.
- **Area Traffic Capacity:** Total traffic throughput served per geographic area.

Moreover, the expected traffic demand is likely to grow steeply with the uptake of NTN ubiquitous connectivity, while current datasets are, by nature, only able to capture and translate past and current user behavioral patterns. Applied in their respective geographical areas, they provide high-level performance targets to influence how the other defined datasets are further processed to translate the expected future user traffic demands more accurately in a 10-year timeline. In a nutshell, the aggregated values acquired from these integrated datasets given in the preceding sections facilitate with the opportunity to indulge into optimizing the multiple resource allocation strategies.

## IV. BEAMFORMING EXAMPLE USING INTEGRATED DATASET

Considering the integration of Ookla and Meta AI datasets for direct-to-smartphone use case, a beamforming example of utilizing the integrated dataset is provided. This example offers initial insights into resource management through NTN system based on user terminal requirements outlined in Table II. The example showcases the projection of a beam through satellite to serve the users in the region where NTN accessibility is truly required. The idea involves initially identifying the regions through Ookla dataset where the average speed falls below 2 Mbps (Megabits per second). The minimum data rate is an essential requirement by user terminal for outdoor scenario as indicated in Table II. Additionally, similar minimum requirements can be considered when terrestrial network accessibility is crumbled under emergency conditions.

In this essence, the 119 tiles from Ookla dataset are identified within the vicinity of Barcelona where the average speed does not meet the requirements. The geocoordinates corresponding to these tiles are then extracted for further processing. Subsequently, the statistical population density is accumulated for the same selected geocoordinates as chosen by using Ookla´s dataset.

Once the total area for those 119 tiles is identified, the beam is projected at the mean point (41.6289 Latitude, 2.2568 Longitude) of the identified region from the simulated satellite. Based on the user terminal pattern experienced in Table II and the number of statistical amount of people, the overall demand at a given time can be estimated roughly and hence, beamforming technique can be applied to project a beam with specific parameters to fulfil the requested throughput. The beam projected on the identified vicinity of Barcelona is depicted in Figure 4.

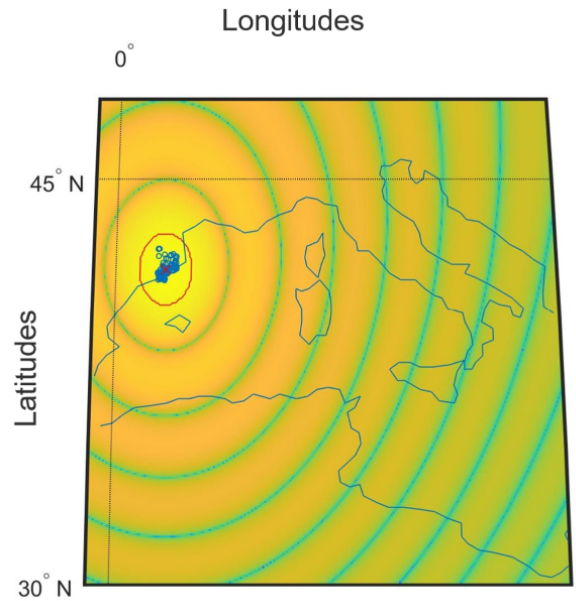

Fig. 4. The beam projected on the identified vicinity of Barcelona.

The red circle is the beam projected by the satellite and blue points are center points of those tiles where the average speed is below 2 Mbps. The accumulated user demand based on the outcomes acquired from the integrated is illustrated in Table III. Note that the demand is roughly estimated to show an example using integrated dataset and the population in percentage needs to be served simultaneously.

TABLE III
ACCUMULATED USER DEMAND BASED ON POPULATION

| No. | Population (Percentage) | Accumulated user demand (Gbps) |
|---|---|---|
| 1 | 10 | 1018 |
| 2 | 7 | 712 |
| 3 | 5 | 509 |
| 4 | 3 | 305 |

The example presented here is on smaller scale, however, for a large scale when there are multiple areas are identified as potential regions for NTN services, an AI based adaptive beamforming technique could be applied. This technique will project the beams based on clustering the areas according to the identified tiles. Since per user requirements will already be known, therefore, the system can be designed to provide sufficient throughput to serve all the users within the identified regions.

## V. CONCLUSIONS

The management of radio resources in NTN is of utmost importance as it directly affects the overall capacity of the communication system. In this context, efficiently allocating resources in a multi-orbital and multi-satellite system becomes challenging due to the dynamic nature of inter and intra satellite links as well as the heterogeneous user traffic demand. To cope with these complexities, there is a growing shift towards leveraging AI for its abilities to handle such problems effectively. Since AI algorithms heavily rely on datasets, this document mainly focused on the identification and the provision of open datasets as potential candidates for AI- enable radio resource controller to effectively manage the satellite resources. In the domain of open datasets, this paper provides access to identified open datasets and offers a comprehensive overview of their relevance for diverse use cases such as maritime, PPDR and direct-to-smartphone etc. Furthermore, it provides valuable guidelines about the localities where NTN is critically required to provide seamless connectivity and reliable communication and then how these datasets can be effectively utilized in managing the radio resources by just taking into the account the given real-world values of the traffic flow and user traffic demand in those localities.

*Data Availability Statement:*
The datasets are openly accessible through the following links https://datasets.cttc.es/ and https://cloud.cttc.es/index.php/s/5YbGTTstnxypabQ

## ACKNOWLEDGEMENT

This activity has been performed in the scope of 6G-NTN project funded by Smart Networks and Services Joint Undertaking (SNS JU) under the European Union's Horizon Research and Innovation program under Grant Agreement No. 101096479 and a part of this work is also supported by the grant from MINECO and EU – NextGenerationEU [UNICO-5G I+D/AROMA3D-SPACE (TSI-063000-2021-70).